\documentclass[12pt]{article}
\usepackage{amssymb,amsmath,epsfig}
\begin{document} \parindent=0pt
\parskip=6pt \rm

 \begin{center}
{\bf \Large Phase transitions to spin-triplet
  ferromagnetic superconductivity in neutron stars}

  \vspace{0.5cm}

{\bf Diana V. Shopova, Tsvetomir E. Tsvetkov, Dimo I. Uzunov},

\vspace{0.5cm}

 {\em CP Laboratory, G. Nadjakov Institute of Solid State Physics,\\ Bulgarian
Academy of Sciences, BG-1784 Sofia, Bulgaria.}
\end{center}

\vspace{0.5cm}

\begin{abstract}
Effects of the anisotropy of Cooper pairs in spin-triplet ferromagnetic
superconductors are investigated on the basis of the Ginzburg-Landau
theory. A special attention is paid to the triggering of the
superconducting state by the ferromagnetic order. The ground states of
these superconductors are outlined and discussed. The idea about a
possible coexistence of ferromagnetism and spin-triplet
superconductivity in neutron stars is introduced.
\end{abstract}

\vspace{0.5cm}

{\bf 1. Introduction}

The cold and dense matter in the interior of neutron stars undergoes
phase transitions to both superfluid and superconducting states (see,
e.g., Refs.~\cite{Tilley:1974, Link:2003, Buckey:2004}. Both the
superfluidity due to neutron Cooper pairing and the superconductivity
due to proton Cooper pairing may be of unconventional spin-triplet
($p$-wave) type~\cite{Tilley:1974}. The latter is known from the theory
of $^3$He liquids~\cite{Leggett:1975,Vollhardt:1990,Volovik:2003} as
well as from the theory of heavy fermion~\cite{Stewart:1984,
Sigrist:1991}) and high-temperature (~\cite{Blagoeva:1990, Uzunov:1990,
Uzunov:1993, Annett:1995}) superconductors.

The possible superconducting phases in unconventional superconductors
are described in the framework of the general Ginzburg-Landau (GL)
effective free energy functional~\cite{Uzunov:1993} with the help of
the  symmetry group theory. Thus a variety of possible superconducting
orderings were predicted for different crystal
structures~\cite{Volovik0:1985, Volovik:1985, Ueda:1985, Blount:1985}.
A detailed thermodynamic analysis~\cite{Blagoeva:1990, Volovik:1985} of
the homogeneous (Meissner) phases and a renormalization group
investigation~\cite{Blagoeva:1990} of the superconducting phase
transition up to the two-loop approximation~\cite{Uzunov:1993} were
also performed.

Recent experiments~\cite{Saxena:2000} at low temperatures ($T \sim 1$
K) and high pressure ($T\sim 1$ GPa) demonstrated the existence of spin
triplet superconducting states in the metallic compound UGe$_2$. This
superconductivity is {\em triggered} by the spontaneous magnetization
($M$) of the ferromagnetic phase ($M$-trigger
effect~\cite{Shopova:2004}). The ferromagnetic order exists at much
higher temperatures and coexists with the superconducting phase in the
whole domain of existence of the latter below $T \sim 1$ K; see also
experiments published in Refs.~\cite{Huxley:2001}. Moreover, the same
phenomenon of existence of superconductivity at low temperatures and
high pressure in the domain of the $(T,P)$ phase diagram where the
ferromagnetic order is present has been observed in ZrZn and URhGe,
too; for details, see, e.g. Ref.~\cite{Shopova1:2003, Shopova2:2003,
Shopova3:2003, Shopova:2004}. Note, that the superconductivity in the
metallic compounds mentioned above, always coexists with the
ferromagnetic order and is enhanced by the latter. Besides, in these
systems the superconductivity seems to arise from the same electrons
that create the band magnetism.

A similar phenomenon of coexistence of ferromagnetism and spin-triplet
superconductivity may exist also in neutron stars. In this case the
Cooper pairing of fermions (protons) will be triggered by the
spontaneous magnetic moment of the same proton subsystem of the nuclear
star matter. The basic features of these phenomena can be described
within an extension of the GL theory. The results can be applied to a
new and interesting problem of coexistence of superconductivity and
ferromagnetism in neutron stars. The coexistence of spin-triplet
superconductivity and ferromagnetism may explain the large magnetic
field of these stars. To our best knowledge the problem of a possible
coexistence of ferromagnetism and superconductivity in neutron stars
and asprophysics objects is introduced for the first time in our
present report.

The theory allows the description of various types of phase transitions
and multicritical points~\cite{Uzunov:1993, Toledano:1987}. Following
notations in Ref.~\cite{Shopova:2004}, here we summarize previous
studies~\cite{Shopova1:2003, Shopova2:2003, Shopova3:2003,
Shopova:2004} and present new results on the effect of the Cooper pair
anisotropy on the phase diagram of spin-triplet ferromagnetic
superconductors~\cite{Shopova:2004}.

Our consideration is focussed on the ground state, namely, we are
interested in uniform phases, where the order parameters (the
superconducting order parameter $\psi$ and the magnetization vector $M
= \left\{M_j, j = 1,2,3\right\}$, do not depend on the spatial vector
$\vec{x}\in V $ ($V$ is the volume of the system).

{\bf 2. Ginzburg-Landau free energy}

Consider the GL free energy $F(\psi,\mbox{\boldmath$M$})=V f(\psi,
\mbox{\boldmath$M$})\:$, where the free energy density
$f(\psi,\mbox{\boldmath$M$})$ (for short hereafter called ``free
energy'') of a spin-triplet ferromagnetic superconductor is given
by~\cite{Shopova:2004}

\begin{eqnarray}
\label{eq2} f(\psi, \mbox{\boldmath$M$}) &= &
 a_s|\psi|^2 +\frac{b_s}{2}|\psi|^4 + \frac{u_s}{2}|\psi^2|^2 +
\frac{v_s}{2}\sum_{j=1}^{3}|\psi_j|^4 +
a_f\mbox{\boldmath$M$}^2 + \frac{b_f}{2}\mbox{\boldmath$M$}^4\\
\nonumber && + i\gamma_0 \mbox{\boldmath$M$}.(\psi\times \psi^*) +
\delta \mbox{\boldmath$M$}^2 |\psi|^2\;.
\end{eqnarray}

In Eq.~(1), $\psi = \left\{\psi_j;j=1,2,3\right\}$ is the
three-dimensional complex vector ($\psi_j = \psi_j^{\prime} + i
\psi_j^{\prime\prime}$) describing the unconventional (spin-triplet)
superconducting order and $\mbox{\boldmath$B$} = (\mbox{\boldmath$H$} +
4\pi\mbox{\boldmath$M$}) = \nabla \times \mbox{\boldmath$A$}$ is the
magnetic induction; $\mbox{\boldmath$H$} = \{H_j; j = 1,2,3\}$ is the
external magnetic field, $\mbox{\boldmath$A$} = \left\{A_j;
j=1,2,3\right\}$ is the magnetic vector potential ($\nabla.\:A = 0$).
In Eq.~(1), $b_s > 0$, $b_f > 0$, $a_f = \alpha_f(T-T_f)$ is given by
the positive material parameter $\alpha_f$ and the ferromagnetic
critical temperature $T_f$ corresponding to a simple superconductor ($M
\equiv 0$), and $a_s = \alpha_s(T-T_s)$, where $\alpha_s$ is another
positive material parameter and $T_s$ is the critical temperature of a
standard second order phase transition which may occur at $|H| =
{\cal{M}} = 0$; ${\cal{M}} = |\mbox{\boldmath$M$}|$. The parameter
$u_s$ describes the anisotropy of the spin-triplet Cooper pair whereas
the crystal anisotropy is described by the parameter
$v_s$~\cite{Blagoeva:1990, Volovik:1985}.

The two orders -- the magnetization vector $M = \left\{M_j\right\}$ and
$\psi = \left\{A_j\right\}$, interact through the last two terms in
(1). The $\gamma_0-$term~\cite{Walker:2002} ensures the triggering of
the superconductivity by the ferromagnetic order ($\gamma_0>0$) whereas
the $\delta-$term makes the model more realistic in the strong coupling
limit~\cite{Machida:2001}. Both $\psi M$-interaction terms included in
(1) are important for a correct description of the temperature-pressure
($T,P$) phase diagram of the ferromagnetic
superconductor~\cite{Shopova:2004}. In general, the parameter $\delta$
for ferromagnetic superconductors may take both positive and negative
values.

The values of the material parameters ($T_s$, $T_f$, $\alpha_s$,
$\alpha_f$, $b_s$, $u_s$, $v_s$, $b_f$, $K_j$, $\gamma_0$ and $\delta$)
depend on the choice of the concrete substance and on intensive
thermodynamic parameters, such as the temperature $T$ and the pressure
$P$. One may assume that the general form (1) of the free energy may
describe the general features of the uniform orders in neutron stars
provided one makes a suitable choice of parameters ($T_s, T_f,
\alpha_s$,...).

As we are interested in the ground state properties, we set the
external magnetic field equal to zero $(H = 0)$. Besides, we emphasize
that the magnetization vector $M$ may produce vortex superconducting
phase~\cite{Tilley:1974, Uzunov:1993} in case of type II
superconductivity. The investigation of nonuniform (vortex) states can
be made with the help of gradient terms in the free
energy~\cite{Shopova:2004} which take into account the spatial
variations of the order parameter field $\psi$. This task is beyond our
present consideration. Rather we investigate the basic problem about
the possible stable uniform (Meissner) superconducting phases which may
coexist with uniform ferromagnetic order. For this aim the free energy
(1) is quite convenient.

In case of a strong easy axis type of magnetic anisotropy, as is in
UGe$_2$~\cite{Saxena:2000}, the overall complexity of mean-field
analysis of the free energy $f(\psi, \mbox{\boldmath$M$})$ can be
avoided by performing an ``Ising-like'' description:
$\mbox{\boldmath$M$} = (0,0,{\cal{M}})$. Further, because of the
equivalence of the ``up'' and ``down'' physical states $(\pm
\mbox{\boldmath$M$})$ the thermodynamic analysis can be performed
within the ``gauge" ${\cal{M}} \geq 0$. But this stage of consideration
can also be achieved without the help of crystal anisotropy arguments.
When the magnetic order has a continuous symmetry one may take
advantage of the symmetry of  the total free energy $f(\psi,
\mbox{\boldmath$M$})$ and avoid the consideration of equivalent
thermodynamic states that occur as a result of the respective symmetry
breaking at the phase transition point but have no effect on
thermodynamics of the system. In the isotropic system one may again
choose a gauge, in which the magnetization vector has the same
direction as  $z$-axis ($|\mbox{\boldmath$M$}| = M_z = {\cal{M}}$) and
this will not influence the generality of thermodynamic analysis. Here
we shall prefer the alternative description within which the
ferromagnetic state may appear through two equivalent ``up'' and
``down'' domains with magnetization $ {\cal{M}}$ and $ -{\cal{M}}$,
respectively.

For our aims we use notations in which the number of
 parameters is reduced. Introducing the parameter
\begin{equation}
\label{eq10} b = (b_s + u_s + v_s) > 0
\end{equation}
we redefine the order parameters and the other parameters in the
following way:
\begin{eqnarray}
\label{eq11} &&\varphi_j =b^{1/4}\psi_j = \phi_je^{\theta_j}\:,\;\;\; M
= b_f^{1/4}{\cal{M}}\:,\\ \nonumber && r =
\frac{a_s}{\sqrt{b}}\:,\;\;\; t =\frac{a_f}{\sqrt{b_f}}\:, \;\;\; w =
\frac{u_s}{b}\:, \;\;\; v =\frac{v_s}{b}\:, \\ \nonumber &&\gamma=
\frac{\gamma_0}{b^{1/2}b_f^{1/4}}\:,\;\;\; \gamma_1=
\frac{\delta}{(bb_f)^{1/2}}\:.
\end{eqnarray}

Having in mind our approximation of uniform $\psi$ and
$\mbox{\boldmath$M$}$
 and the notations~(2)~-~(3), the free energy density
 $f(\psi,M)$ can be written in the form
\begin{eqnarray}
\label{eq12} f(\psi,M)& = & r\phi^2 + \frac{1}{2}\phi^4
  + 2\gamma\phi_1\phi_2 M \mbox{sin}(\theta_2-\theta_1) + \gamma_1 \phi^2 M^2
+ tM^2 + \frac{1}{2}M^4\\ \nonumber && -2w
\left[\phi_1^2\phi_2^2\mbox{sin}^2(\theta_2-\theta_1)
 +\phi_1^2\phi_3^2\mbox{sin}^2(\theta_1-\theta_3) +
 \phi_2^2\phi_3^2\mbox{sin}^2(\theta_2-\theta_3)\right] \\ \nonumber
&& -v[\phi_1^2\phi_2^2 + \phi_1^2\phi_3^2 + \phi_2^2\phi_3^2].
\end{eqnarray}
In this free energy the order parameters $\psi$ and
$\mbox{\boldmath$M$}$ are defined per unit volume.

In contrast to the situation in superconducting compounds, for the case
of neutron stars, the crystal field anisotropy represented by the
$v_s-$terms in (1)~--~(4) can be ignored, and for this reason we shall
discuss the case $v_s \equiv 0$. We assume that $T_f > T_s$. This is
the case when the superconductivity is triggered by the magnetic order.
Besides we shall discuss the stable phases in the temperature region $T
> T_s$. The case $T_f < T_s$ may also present interest for neutron
stars and, hence, it needs a special investigation. As mentioned in
Ref.~\cite{Shopova:2004}, the case $T_s \sim T_f$ allows for a quite
simple analytical treatment. All these cases may be of interest to the
description of ferromagnetic superconductivity in stellar objects
whereas in condensed matter only cases of $T_f \gg T_s$ have been
observed so far.

Our consideration is performed within the framework of the standard
mean-field analysis~\cite{Uzunov:1993}. The stable phases correspond to
global minima of the GL energy (1). The equilibrium phase transition
line separating two phases is defined by the thermodynamic states,
where the respective GL free energies are equal.

{\bf 3. Phases and phase diagram}

 The calculations show that for temperatures $T > T_s$, i.e., for $r > 0$,
we have three stable phases. Two of them are quite simple: the normal
($N$-) phase ($\psi = M = 0$) with existence and stability domains
given by $t>0$ and $r>0$, and the ferromagnetic phase (FM) given by
$\psi =0$ and $M^2 = -t$ which has the existence condition $ t<0$, and
a stability domain defined by the inequalities $r
> \gamma_1t$ and
\begin{equation}
\label{eq39} r > \gamma_1t + \gamma\sqrt{-t}\:.
\end{equation}
The third stable phase is a phase of coexistence of superconductivity
and ferromagnetism (hereafter referred to as FS). It is given the
following equations:
\begin{equation}
\label{eq40} \phi_1 = \phi_2=\frac{\phi}{\sqrt{2}}\:, \;\;\; \phi_3 =
0\:,
\end{equation}
\begin{equation}
\label{eq41} \phi^2= (\pm \gamma M-r-\gamma_1 M^2)\:,
\end{equation}
\begin{equation}
\label{eq42} (1-\gamma_1^2)M^3\pm \frac{3}{2} \gamma \gamma_1 M^2
+\left(t-\frac{\gamma^2}{2}-\gamma_1 r\right)M \pm \frac{\gamma
r}{2}=0\:,
\end{equation}
and
\begin{equation}
\label{eq43} (\theta_2 - \theta_1) = \mp \frac{\pi}{2} + 2\pi k\:,
\end{equation}
($k = 0, \pm 1,...$). The upper sign in Eqs.~(7)~--~(9) corresponds to
a domain in which $\mbox{sin}(\theta_2-\theta_1) = -1$ and the lower
sign corresponds to a second domain which may be referred to as
FS$^{*}$; in the latter, $\mbox{sin}(\theta_2-\theta_1) = 1$. These two
domains are equivalent and describe the same ordering. We shall focus
on the upper sign in (7)~--~(9), i.e. on FS.

The phase diagram ($t,r$) is outlined in Figs.~1 and 2 for different
values of the anisotropy parameter $w$. The phase transition lines for
$w > 0$ and $w < 0$ shown in Figs.~1 and 2, respectively, have
qualitatively the same shape as the phase transition lines
corresponding to $w = 0$~\cite{Shopova:2004} but there are essential
quantitative differences between these cases. We shall discuss them in
the next section.

\begin{figure}
\begin{center}
\epsfig{file=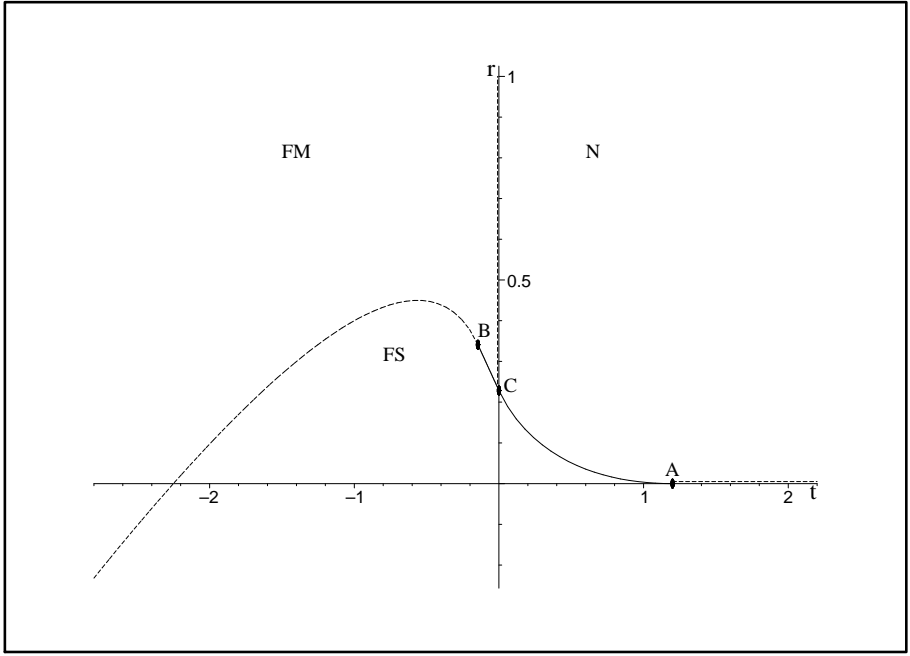,angle=-90, width=12cm}
\end{center}
\caption{Phase diagram in the ($t$, $r$) plane for $\gamma = 1.2, $
$\gamma_1 = 0.8$, and $w = 0.4$.} \label{STUf1.fig}
\end{figure}
\begin{figure}
\begin{center}
\epsfig{file=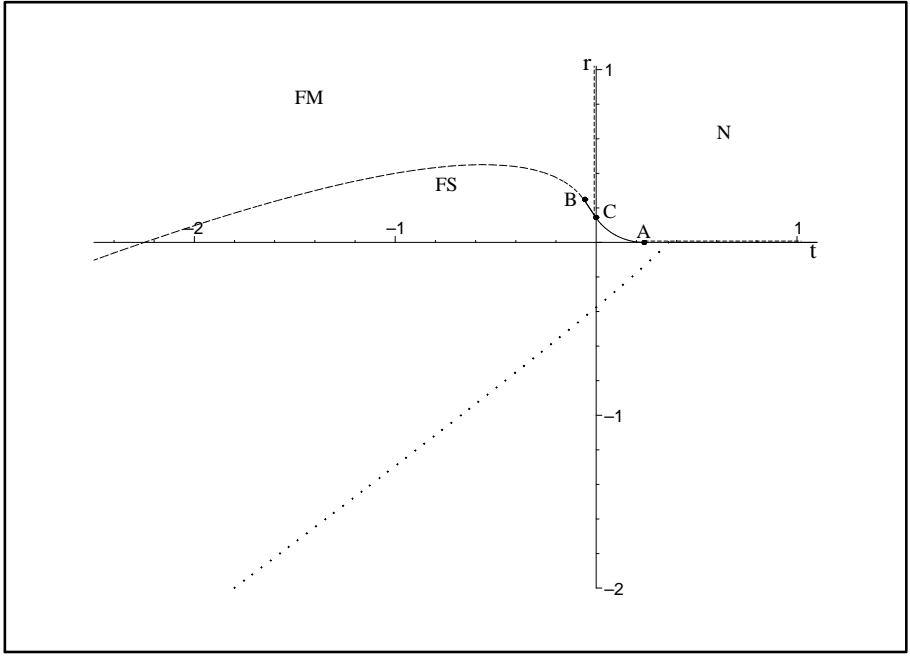,angle=-90, width=12cm}
\end{center}
\caption{Phase diagram in the ($t$, $r$) plane for $\gamma = 1.2, $
$\gamma_1 = 0.8$, and $w = -2$. The various lines are explained in the
text.} \label{STUf2.fig}
\end{figure}

Note, that the phase diagrams in Figs.~1 and 2 exhibit two types of
phase transitions. The dashed curves indicate second order phase
transitions of type N-FM, FM-FS and N-FS, whereas the curve AC and the
straight line BC indicate the first order phase transitions N-FS and
FM-FS, respectively. The points $A$ and $B$ are tricritical whereas $C$
is a triple point, where N, FM and FS coexist~\cite{Uzunov:1993}. The
negative values of the parameter $r$ are restricted by $r(T) \geq
r(T=0)$. Having in mind this condition as well as the shape of the
phase diagram (Figs.~1 and 2) we easily conclude that both FM and FS
ground states (at $T =0$) are possible in systems described by the
model (1). This is the case for the itinerant magnets, mentioned above,
and one may speculate that this situation may occur in neutron stars,
too. Whether FM and FS ground state will occur depends on the
particular values of the material parameters ($T_s$, $T_f$,
$\alpha_s,...$).

The final aim of the phase diagram investigation is the outline of the
($T,P$) diagram. Important conclusions about the shape of the $(T,P)$
diagram can be made from the form of the $(t,r)$ diagram without an
additional information about the values of the relevant material
parameters $(a_s$, $a_f,...$) and their dependence on the pressure $P$.
For example, the equilibrium temperature $T_{FS}$ of the phase
transitions to FS phase varies with the variation of the system
parameters $(\alpha_s, \alpha_f,...)$ from values which are much higher
than the characteristic temperature $T_s$ up to zero temperature.

{\bf 4. Anisotropy effects}

Our analysis demonstrates that when the anisotropy of the Cooper pairs
is taken in consideration, there will be not drastic changes in the
shape the phase diagram for $r>0$ and the order of the respective phase
transitions. Of course, there will be some changes in the size of the
phase domains and the formulae for the thermodynamic quantities. This
is seen from Figs.~1 and 2 which are shown for the first time in the
present report. Besides, it is readily seen from Figs.~1 and 2 that the
temperature domain of first order phase transitions and the temperature
domain of stability of FS above $T_s$ essentially vary with the
variations of the anisotropy parameter $w$. The parameter $w$ will also
insert changes in the values of the thermodynamic quantities like the
magnetic susceptibility and the entropy and specific heat jumps at the
phase transition points~\cite{Shopova3:2003}.

Besides, and this seems to be the main anisotropy effect, the $w$- and
$v$-terms in the free energy lead to a stabilization of the order along
the main crystal directions which, in other words, means that the
degeneration of the possible ground states is considerably reduced.
This means also a smaller number of marginally stable states.

The dimensionless anisotropy parameter $w$ can be either positive or
negative depending on the sign of $u_s$. Obviously when $ u_s > 0$, the
parameter $w$ will be positive too ($0< w<1$). We shall illustrate the
influence of Cooper-pair anisotropy in this case. The order parameters
($M$, $\phi_j$, $\theta_j$) are given by Eqs.~(6), (9),
 \begin{equation}
\label{eq54} \phi^2=\frac{\pm \gamma M-r-\gamma_1 M^2}{(1-w)} \ge 0\:,
\end{equation}
and
\begin{equation}
\label{eq55} (1- w - \gamma_1^2)M^3 \pm \frac{3}{2} \gamma \gamma_1 M^2
+\left[t(1-w)-\frac{\gamma^2}{2}-\gamma_1 r\right]M \pm \frac{\gamma
r}{2}=0\:,
\end{equation}
where the meaning of the upper and lower sign is the same as explained
just below Eq.~(9). We consider the FS domain corresponding to the
upper sign in the Eq.~(10) and (11). The stability conditions for FS
read,
\begin{equation}
\label{eq56} \frac{ (2-w)\gamma M- r -\gamma_1M^2}{1-w} \ge 0\:,
\end{equation}
\begin{equation}
\label{eq57} \gamma M -wr-w \gamma_1 M^2 > 0\:,
\end{equation}
and
\begin{equation}
\label{eq58} \frac{1}{1-w}\left[3(1-w-\gamma_1^2) M^2 + 3 \gamma
\gamma_1 M + t(1-w)-\frac{\gamma^2}{2} -\gamma_1 r \right]\geq 0\:.
\end{equation}
For $M\ne (\gamma/2 \gamma_1)$ we can express the function $r(M)$
defined by Eq.~(10), substitute the obtained expression for $r(M)$ in
the existence and stability conditions (10)-(14) and do the analysis in
the same way as for $w=0$~\cite{Shopova:2004}. The most substantial
qualitative difference between the cases $w>0$ and $w<0$ is that for $w
< 0$ the stability of FS is limited for $r<0$. This is seen from Fig.~2
where FS is stable above the straight dotted line for $r < 0$ and $t <
0$. This includes into consideration also purely superconducting
(Meissner) phases as ground states.

{\bf 5. Final remarks}

We have done an investigation of the M-trigger effect in unconventional
ferromagnetic superconductors. This effect due to the
$M\psi_1\psi_2$-coupling term in the GL free energy consists of
bringing into existence of superconductivity in a domain of the phase
diagram of the system that is entirely in the region of existence of
the ferromagnetic phase. This form of coexistence of unconventional
superconductivity and ferromagnetic order is possible for temperatures
above and below the critical temperature $T_s$, which corresponds to
the standard phase transition of second order from normal to Meissner
phase -- usual uniform superconductivity in a zero external magnetic
field, which appears outside the domain of existence of ferromagnetic
order. Our investigation has been mainly intended to clarify the
thermodynamic behaviour at temperatures $T_s< T < T_f$, where the
superconductivity cannot appear without the mechanism of M-triggering.
We have described the possible ordered phases (FM and FS) in this most
interesting temperature interval.

The Cooper pair and crystal anisotropies have also been investigated
and their main effects on the thermodynamics of the triggered phase of
coexistence have been established. In discussions of concrete real
material one should take into account the respective crystal symmetry
but the variation of the essential thermodynamic properties with the
change of the type of this symmetry is not substantial when the low
symmetry and low order (in both  $M$ and $\psi$) $\gamma$-term is
present in the free energy.

Below the superconducting critical temperature $T_s$ a variety of pure
superconducting and mixed phases of coexistence of superconductivity
and ferromagnetism exists and the thermodynamic behavior at these
relatively low temperatures is more complex than in known cases of
improper ferroelectrics; see. e.g., Ref.~\cite{Toledano:1987}. The case
$T_f < T_s$ also needs a special investigation.

Our results are referred to the possible uniform superconducting and
ferromagnetic states. Vortex and other nonuniform phases need a
separate study.

More experimental information about the values of the material
parameters ($a_s, a_f, ...$) included in the free energy (1) is
required in order to outline the thermodynamic behavior and the phase
diagram in terms of thermodynamic parameters $T$ and $P$. In
particular, a reliable knowledge about the dependence of the parameters
$a_s$ and $a_f$ on the pressure $P$, the value of the characteristic
temperature $T_s$ and the ratio $a_s/a_f$ at zero temperature are of
primary interest.

The phenomenological GL model (1) is quite general and can be reliably
used in considerations of a possible coexistance of ferromagnetism and
unconventional superconductivity in stellar objects. Recent
investigations~\cite{Link:2003, Buckey:2004} of superconductivity in
neutron stars can be related with the present consideration.

{\bf Acknowledgments:}

 One of us (T.E.T.) thanks the hospitality of Dr. V. Celebonovic and the
 Organizers of the Workshop on Equation of State and Phase Transition Issues in
 Models of Ordinary Astrophysical Matter (Lorentz Center, Leiden University, 2-11 June 2004).
  Financial support by SCENET (Parma) and JINR (Dubna) is also
acknowledged.

\end{document}